\documentclass[aps,showpacs,pra]{revtex4}
\usepackage{amsmath}
\usepackage{amsfonts}
\usepackage{amssymb}
\usepackage{epstopdf}
\usepackage{graphicx}
\begin{document}
\title{{Adiabatic Generation of N-quNit Singlet States with Cavity QED}}
\author{Li-Xiang Ye$^{1,2}$, Xiu Lin$^{1,2}$, and Rong-Can Yang$^{1,2}$}
\email{rcyang@fjnu.edu.cn}
\author{Hong-Yu Liu$^{3}$}
\affiliation{$^1$ Fujian Provincial Key Laboratory of Quantum Manipulation and New Energy Materials,\\ College of Physics and Energy, Fujian Normal University, Fuzhou, 350007, China}
\affiliation{$^2$ Fujian Provincial Collaborative Innovation Center for Optoelectronic Semiconductors and Efficient Devices, Xiamen 361005, China}
\affiliation{$^3$ College of Science, Yanbian University, Yanji, 133002, China}

\date{\today}
\begin{abstract}
A theoretical scheme is presented for the adiabatic generation of N-quNit singlet states with $N\geqslant3$, which may be more feasible than previous ones in a cavity.
In this proposal, the system may be robust both parameter fluctuations and dissipation along a dark state.
In addition, quantum information is only stored in atomic ground states and there is no energy exchanged between atoms and photons in a cavity so as to reduce the influence of atomic spontaneous emission and cavity decays.
\end{abstract}
\maketitle

\section{Introduction}

As a powerful tool in quantum theory, quantum entanglement can be employed not only to test quantum nonlocality \cite{Kaszlikowski}, but also have been shown to various applications in quantum information processing \cite{Bennett,Cirac,Mattle,Vidal,Hillery,Wiesner}.
It is well known that high-dimensional states have more specific properties than bipartite ones.
An N-quNit singlet state with its total spin being zero is an example, which was proposed by Cabello in 2002 \cite{Cabello}.
Its mathematical expression is
\begin{equation}\label{sn}
|S_N^{(N)}\rangle = \frac{1}{{\sqrt {N!} }}\sum\limits_{\scriptstyle permutations\hfill\atop
\scriptstyle of\,01 \cdots \left( {N - 1} \right)\hfill} {{{\left( { - 1} \right)}^t}\left| {ij \cdots n} \right\rangle },
\end{equation}
where $t$ is the number of transpositions of pairs of elements that must be composed to place the elements in canonical order.
These entangled states can be used in proofs of Bell's theorem without inequalities \cite{Mermin} and play an important role in constructing decoherence-free subspaces \cite{Zanardi}.In addition, singlet states can also be applied to realize an unknown unitary transformation \cite{MHillery} and solve several practical problems without classical solutions, including ``N -strangers'', ``secret sharing'' and ``liar detection'' \cite{ACabello}.

Although a singlet state possesses lots of distinctive features, Cabello had ever pointed that it was a great challenge to prepare this kind of states with $N\geqslant3$.
So far, there have been several scenarios proposed to generate singlet states involving three or more atoms, but most of them only refer to three-qutrit singlet states, in which two main methods are used \cite{Jin,Lin,Huang,Yang,Shao,Zheng,Shi,You,RCYang,LXYe}.
One is to need several steps which was first raised by Jin et al.\cite{Jin}, while the other one is to add a separated state to a lower-dimensional singlet state in one step, which was first put forward by our group in 2009 \cite{Yang}.
Obviously, the second one may be a potential candidate to realize any-dimension singlet states.
In 2010, Shao et al. used the mode of $\Lambda$-like multilevels atoms interacting with a multi-mode cavity with similar ways \cite{Shao}.Although the idea is exciting, it may be hardly realized for high-dimension ones with $N>3$.
Because of it, we propose an alternative scheme with the combination of Rydberg blockade and adiabatic passage technologies in 2016 \cite{XLin}, which may provide a new idea for the preparation of N-quNit singlet states and other high-dimension entangled states.
Of course, we should admit that the proposal may be very difficult due to the utilization of ladder-type Rydberg atoms.

In this paper, we propose another scheme to generate N-quNit singlet states in a cavity.
Different from previous ones, we store quantum information in ground states of wave-type atoms and use a single-mode cavity to act as a medium, which has been broadly used before\cite{Raimond}.
Thus, our protocol would be feasible.

The paper is organized as follows.
In Sec. II, details for the generation of N-quNit singlet states are described.
Sec. III makes numerical simulations and analysis.
A conclusion is made in Sec. IV.

\section{Generation of N-atom singlet states}

\begin{figure}
  \includegraphics[width=10cm]{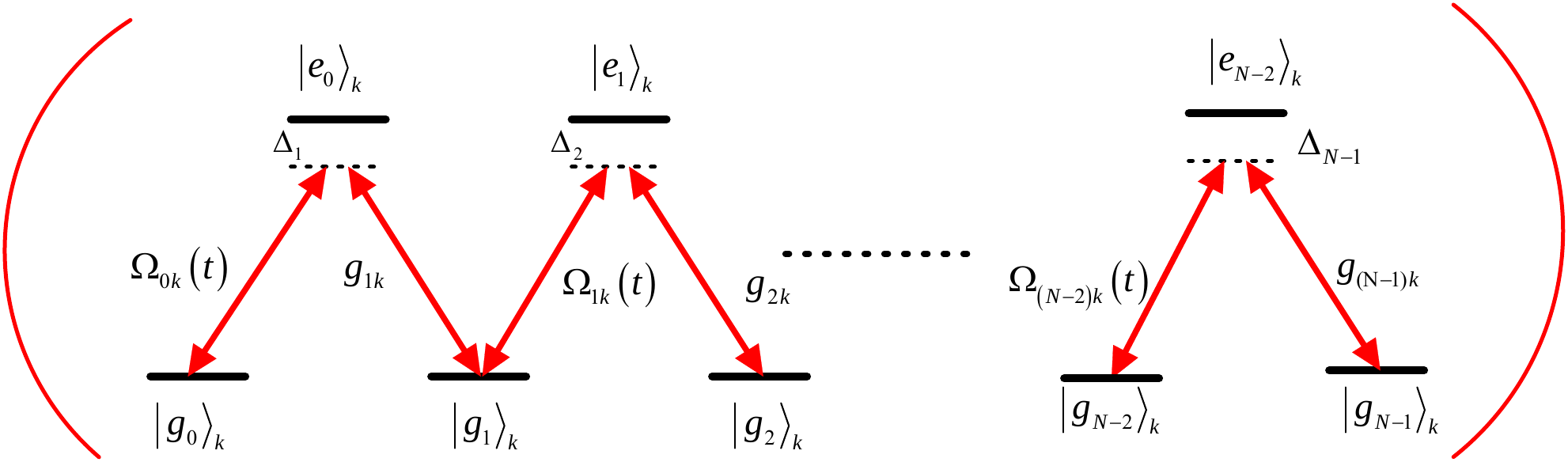}\\
  \caption{(Color online)Atomic level configurations}\label{1ct}
\end{figure}

As illustrated in Fig.\ref{1ct}, we consider that $N$ atoms are trapped in a cavity, where each atom has $N$ ground states ${|g_0\rangle}_k,{|g_1\rangle}_k,..., {|g_{N-1}\rangle}_k (k=1, 2,..., N)$ and $N-1$ excited states ${|e_0\rangle}_k,{|e_1\rangle}_k,..., {|e_{N-2}\rangle}_k$ with the subscript $k$ representing the $k$th atom or a laser pulse (cavity mode) interacting with the $k$th atom in the whole paper.
The atomic transition ${|g_j\rangle}_k \leftrightarrow {|e_j\rangle}_k (j=0,1,..., N-2)$ is driven by a classical laser pulse with time-dependent Rabi frequencies $\Omega_{jk}(t)$ and detuning $\Delta_{j+1}$, while the transition ${|e_j\rangle}_k \leftrightarrow {|g_{j+1}\rangle}_k$ is coupled to the cavity mode with coupling constant $g_{j+1,k}$ and the same detuning $\Delta_{j+1}$.
Then, with the consideration of resolved sidebands and rotating-wave approximations, the total Hamiltonian for this system, in the interaction picture, reads $(\hbar=1)$
\begin{equation}\label{HI}
\begin{split}
H_I&=\sum\limits_{k=1}^N \sum\limits_{j=0}^{N-2}\left(\Omega_{jk}{|e_j\rangle}_k \langle g_j| e^{i\Delta_{j+1}t}\right.\\
&\left.+g_{j+1,k}{|g_{j+1}\rangle}_k \langle e_j| a^+ e^{-i\Delta_{j+1}t}\right)+H.c.,
\end{split}
\end{equation}
where $a^+$ is a creation operator of the cavity. Taking account of large detuning conditions, i.e. $\Delta_{p+1}>>\Omega_{jk}, g_{j+1,k} (p,j=0,1,...,N-2)$, all excited states ${|e_j\rangle}_k$  can be adiabatically eliminated, leading to the effective Hamiltonian
\begin{equation}\label{He1}
\begin{split}
{H_e}&=-i {H_I}\int{{H_I} dt}\\
     &=\sum\limits_{k=1}^N \sum\limits_{j=0}^{N-2}\left[A_{jk}{|g_j\rangle}_k \langle g_j| + G_{jk}{|g_{j+1}\rangle}_k \langle g_{j+1}| a^+ a \right.\\
     &\left.+\left(B_{jk}{|g_{j+1}\rangle}_k \langle g_j| a^+ + H.c. \right) \right],
\end{split}
\end{equation}
where $A_{jk}=-\frac{\Omega_{jk}^2}{\Delta_{j+1}}$,$G_{jk}=-\frac{g_{j+1,k}^2}{\Delta_{j+1}}$, $B_{jk}=-\frac{\Omega_{jk} g_{j+1,k}}{\Delta_{j+1}}$, $(j=0,1,...,N-2; k=1,2,..., N)$.
The first term describes Stark shift induced by a classical laser, the second one shows photo-number-dependent Stark shift and the third denotes the interaction between cavity mode with the transition $|g_{j+1}\rangle \leftrightarrow |g_j\rangle$, respectively.
By using the nonresonant coupling of other lasers with the corresponding atom levels, these energy level shifts can be compensated straightforwardly \cite{Pellizzari}, then we have
\begin{equation}\label{He2}
\begin{split}
H_e &= \sum\limits_{k=1}^N \sum\limits_{j=0}^{N-2}\left[G_{jk}{|g_{j+1}\rangle}_k \langle g_{j+1}| a^+ a \right.\\
    &\left.+ \left(B_{jk}{|g_{j+1}\rangle}_k \langle g_j| a^+ +H.c.\right) \right].
\end{split}
\end{equation}

\begin{figure}
  \includegraphics[width=10cm]{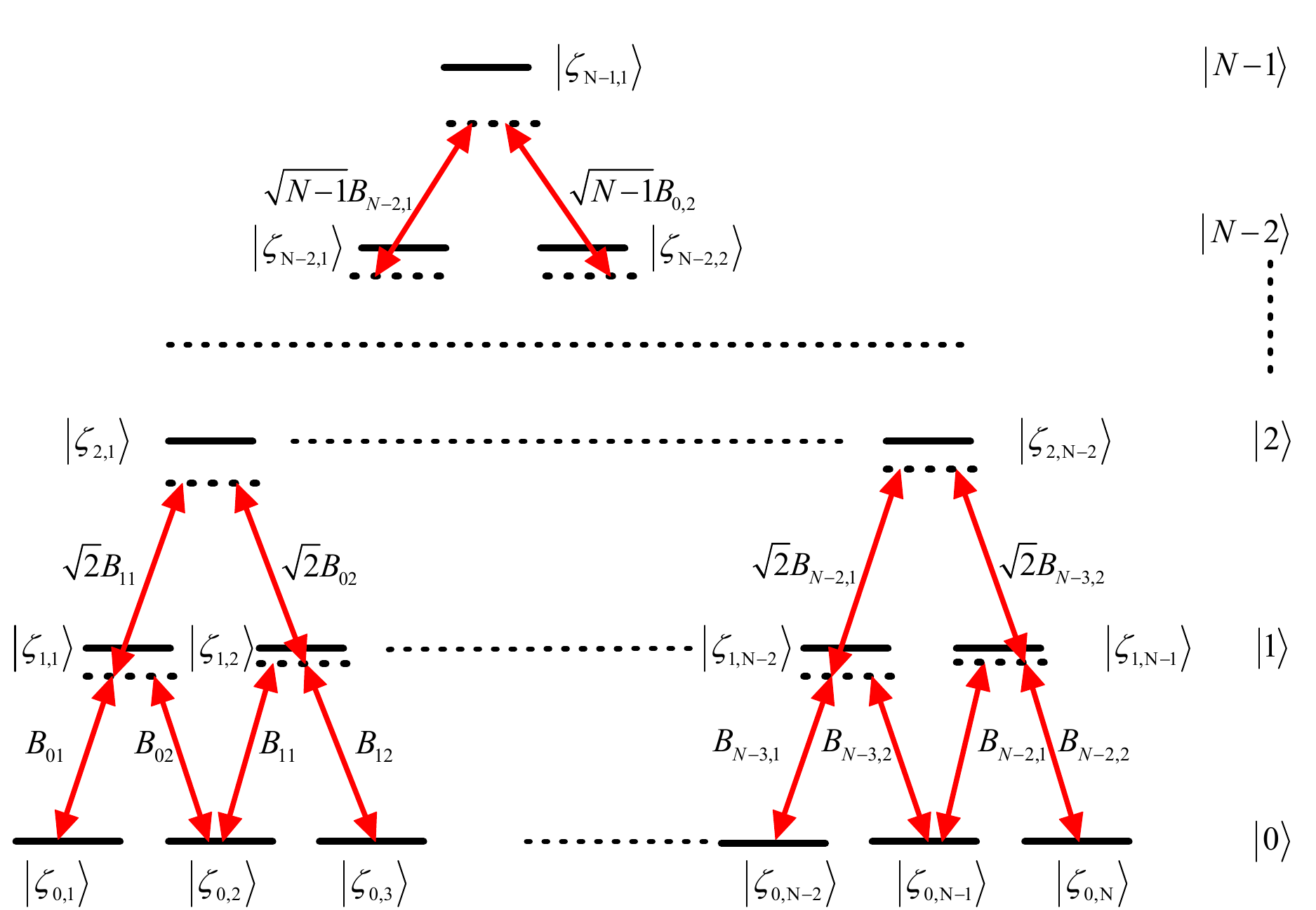}\\
  \caption{(Color online)Logic states and transitions for the generation of N-atom singlet states.}\label{2ct}
\end{figure}

For the sake of simplicity, all atoms except the first one are assumed to have same parameters, i.e. $X_{jk}=X_{j2} (X=\Omega,g,\Delta, A, B, G; k=2,3,..., N)$  and Stark shifts for all levels are set to be equal, i.e. $G_{0k}=G_{1k}=...=G_{(N-2)k}=G$ $(k=1,2,...,N)$.
It should be pointed out although Stark shifts $G_{jk}$ for all levels can not directly be equal, but we can use compensating technologies to let them be approximately equal, which is sufficient for our proposal.
Further, if the initial system state is $|\zeta_{0,N}\rangle = |g_{N-1}\rangle |S_{N-1,0}\rangle |0\rangle$ with the first atom initially being in $|g_0\rangle$, atoms from $2$ to $N$ preparing in a singlet state $|S_{N-1,N-1}\rangle$, and the cavity in the vacuum state $|0\rangle$, then the whole system will be confined in the closed subspace $\{|\zeta_{j,m}\rangle\}$ $(0 \leqslant j \leqslant N-1, 1 \leqslant m \leqslant N-j)$ without the consideration of any dissipations, where
\begin{equation}\label{Basis}
\begin{split}
|\zeta_{j,m}\rangle &= |g_{j+m-1}\rangle |S_{N-1,N-m}\rangle |j\rangle,\\
\left| {{S_{N - 1,0}}} \right\rangle  &= \frac{1}{{\sqrt {\left( {N - 1} \right)!} }}\sum\limits_{\scriptstyle permulations\,\;of\hfill\atop
\scriptstyle{g_0},\,{g_1},\, \cdots ,{g_{N - 2}}\hfill} {{{\left( { - 1} \right)}^t}\left| {i,j, \cdots ,n} \right\rangle } ,\\
|S_{N-1,q+1}\rangle &= {\sigma}^+ |S_{N-1,q}\rangle = \left(|g_1\rangle \langle g_0| + |g_2\rangle \langle g_1| \right.\\
                    &\left.+ ... + |g_{N-1}\rangle \langle g_{N-2}|\right)|S_{N-1,q}\rangle (q \geqslant 0).
\end{split}
\end{equation}
In this case, the Hamiltonian for the whole system reduces to
\begin{equation}\label{HW}
\begin{split}
H&=GN \sum\limits_{j=0}^{N-1}{j|\zeta_{j,1}\rangle \langle \zeta_{j,1}|} + G(N-1)\sum\limits_{j=0}^{N-1}\sum\limits_{m=2}^{N-j}{j|\zeta_{j,m}\rangle \langle \zeta_{j,m}|}\\
 &+\sum\limits_{j=0}^{N-1}\sum\limits_{m=1}^{N-1-j}{}\left(\sqrt{j+1}B_{j+m-1,1}|\zeta_{j,m}\rangle \langle \zeta_{j+1,m}|\right.\\
 &\left.+\sqrt{j+1}B_{j+m-1,1}|\zeta_{j,m+1}\rangle \langle \zeta_{j+1,m}|\right)\\
\end{split}
\end{equation}
To demonstrate this Hamiltonian more clearly, we depict the relation of $|\chi_{jm}\rangle$ with different $j$ and $m$ in Fig.\ref{2ct}, where, for example, the lowest levels correspond to the system state with the cavity mode in the vacuum states.
Through solving the eigenvalue equation of $H$, we obtain a time-dependent dark state with a null eigenvalue
\begin{equation}\label{Dt}
\begin{split}
\left| {D\left( t \right)} \right\rangle &\sim \sum\limits_{p = 0}^{N - 1} {{\left( { - 1} \right)}^p}\left( {\prod\limits_{j = 0,p \leqslant p - 1}^{p - 1} {{B_{j,1}}} \prod\limits_{m = p,m \leqslant N - 2}^{N - 2} {{B_{m,2}}} } \right) \\
&\times\left|{{\zeta _{0,p + 1}}} \right\rangle
\end{split}
\end{equation}

For simplicity, we set $B_{jk} = \chi_j B_{0k}(t)$ $(k=1,2)$ with $\chi_j$ being time-independent. Due to $B_{jk}=-\frac{\Omega_{jk}g_{j+1,k}}{\Delta_{j+1}}$, if we design laser pulses satisfying
\begin{equation}\label{tlim}
\mathop {\lim }\limits_{t \to  - \infty } \frac{{{B_{02}}\left( t \right)}}{{{B_{01}}\left( t \right)}} = 0,\;\mathop {\lim }\limits_{t \to  + \infty } \frac{{{B_{02}}\left( t \right)}}{{{B_{01}}\left( t \right)}} = 1,\;
\end{equation}
then we can adiabatically transfer the initial system state $|\zeta_{0,N}\rangle$ to an N-atom singlet state
\begin{equation}\label{DF}
|D(+\infty)\rangle \sim \frac{1}{\sqrt{N!}}\sum\limits_{k=0}^N {(-1)^k |\zeta_{0,k+1}\rangle}=|S_N\rangle.
\end{equation}

\section{Numerical analysis}

According to Eq.(\ref{tlim}), we can design two time-dependent pulses to be \cite{Vitanov}
\begin{equation}\label{omega}
\begin{split}
\Omega_{01} &= \Omega_0 \exp\left(-\frac{(t-\tau)^2}{T^2}\right)+\Omega_0 \exp \left(-\frac{(t+\tau)^2}{T^2}\right),\\
\Omega_{02} &= \Omega_0 \exp\left(-\frac{(t-\tau)^2}{T^2}\right),
\end{split}
\end{equation}
where $\Omega_0$ represents the time-independent amplitude and $T$ ($\tau$) the pulse width (delay).

\begin{figure}
   \setlength{\abovecaptionskip}{-3cm} 
   \setlength{\belowcaptionskip}{-0cm} 
  \includegraphics[width=10cm]{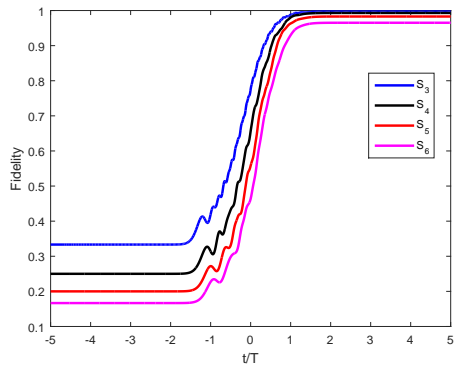}\\
  \caption{(Color online)An explicit example for the fidelity of the generated singlet states for $N=3,4,5,6$ varying with the interaction time $t$, where $T=800/\Omega_0, \tau=T/2, g=\Omega_0, \Delta=10\Omega_0$ have been chosen}\label{3ct}
\end{figure}

To begin with, the procedure of the generation of singlet states is illustrated.
In order to do it, we define the fidelity $F_N=|\langle S_N| \rho (t)|S_N\rangle|$ for the singlet state with $N$ atoms with $\ddot{\rho}(t)$ satisfying the Schr$\ddot{o}$dinger equation $\dot{\rho}=-i[H_e,\rho]$.
In Fig.\ref{3ct}, we only give the illustration for $N=3, 4, 5, 6$ with the choice of $T=\frac{800}{\Omega_0}, \tau=\frac{T}{2}, g=\Omega_0, \Delta=10\Omega_0$, which can also show the similar procedure for $N\geqslant7$.
Results show that the fidelity for $N=3$ to $5$ can rise to the value close to $0.99$, and the maximum fidelity for $N=6$ can also reach about 0.97, meaning that a singlet state for three, four, five, or six atoms can be generated.
In addition, we can predict a singlet state for $N$ atoms with $N\geqslant7$ can be realized if we have prepared the singlet state with $N-1$ atoms.

\begin{figure}
   \setlength{\abovecaptionskip}{-3cm} 
   \setlength{\belowcaptionskip}{-0cm} 
  \includegraphics[width=10cm]{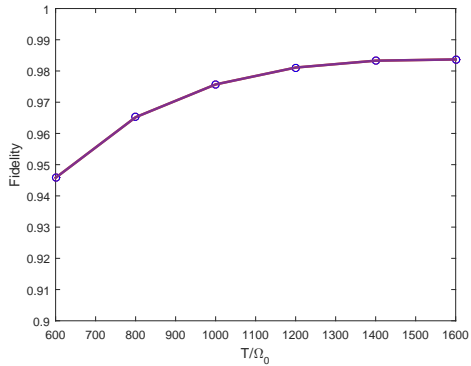}\\
  \caption{(Color online)Fidelity of the prepared states versus pulse width, where parameters $g=\Omega_0, \Delta=10\Omega_0, \tau=T/2, t=5T$}\label{4ct}
\end{figure}

However, we should admit that the fidelity will decrease quickly with the increase of $N$ when all parameters have been set.
For example, it is noticed that the fidelity of the generated singlet states for $5$ atoms is about 0.99, whereas the fidelity for $6$ atoms is only 0.97 in Fig.\ref{3ct}.
How can we improve the final fidelity for $6$ atoms?
In this paper, we raise the pulse width, pulse delay and the required time to increase the fidelity.
In Fig.\ref{4ct}, the graph illustrates the relation between the fidelity for $6$ atoms and pulse width, where we have set $g=\Omega_0, \Delta=10\Omega_0, \tau=\frac{T}{2}, t=5T$.
It is clearly demonstrated that the larger $T$ is, the greater the fidelity $F$ is, and the fidelity will finally reach a stable value which is close to 1.
For instance, the fidelity raises from $F=0.946$ to $0.981$ with the increase of $T$ from $\frac{600}{\Omega_0}$ to $\frac{1200}{\Omega_0}$.
Thus, in ideal conditions, we can enlarge corresponding parameters to improve the fidelity.
But, we should also accept the fact that it may result in some other problems, such as the effect of photon decay and spontaneous emission.
Therefore, our protocol may only be realized for limited number of atoms.
In the following paragraphs, we focus on laser pulse width $T=\frac{800}{\Omega_0}$.

\begin{figure}
   \setlength{\abovecaptionskip}{-3cm} 
   \setlength{\belowcaptionskip}{-0cm} 
  \includegraphics[width=10cm]{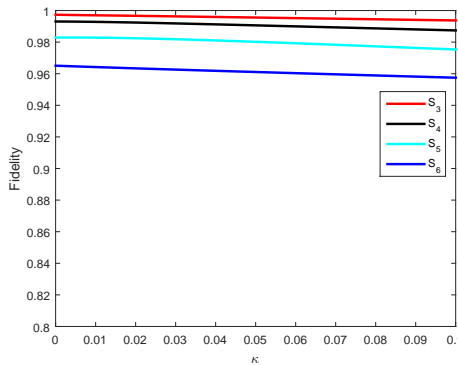}\\
  \caption{(Color online)Fidelity of the prepared states versus the cavity decay reate $\kappa$, where parameters $g=\Omega_0, \Delta=10\Omega_0, T=800/\Omega_0,\tau=T/2, t=5T$}\label{5ct}
\end{figure}

Last but not the least, we take the effect of photon decay into account.
To achieve this motivate, we resort to the master equation expressed as
\begin{equation}
\dot{\rho}=-i[H_e,\rho]-\frac{\kappa}{2}(a^+ a\rho - 2a \rho a^+ + \rho a^+ a)
\end{equation}
where $\kappa$ denotes the decay rate of the cavity mode.
In Fig.\ref{5ct}, we only depict the curve for the relationship of the fidelity versus the decay for $3\leqslant N\leqslant 6$, where we have set $g=\Omega_0, \Delta=\Omega_0, T=\frac{800}{\Omega_0}, \tau=\frac{T}{2}, t=5T$.
From the graph, we can clearly see that the fidelity of the generated singlet states is scarcely insensitive to cavity decay.
For example, the fidelity for $N=3$ only drops from $F=0.997$ to $0.994$ when the cavity decay rate increases from $\kappa=0$ to $0.1$, while the fidelity for $N=6$ only drops from $F=0.965$ to $0.957$.
It should be noticed that the system must be influenced by spontaneous emission of atoms, But this effect may be suppressed with all atoms only virtually excited.

\section{Feasibility and conclusions}

In this part, we first make a discussion on the feasibility.
The atomic levels used in our proposal can easily be found in hyperfine states of natural or artificial atoms, such as Cs atoms.
Of course, we should admit that the scale for the generated singlet states is limited by atomic structures.
For example, if we choose hyperfine states $|F=4, m_F=-4, -3, ..., 4\rangle$ of $6^2S_{1/2}$ of $^{133}Cs$ atoms to act as ground states $|g_j\rangle (j=0, 1, ..., 8)$ and $|F=5, m_F=-4, -3, ..., 3\rangle$ of $6^2P_{3/2}$ to act as excited states $|e_j\rangle (j=0, 1, ..., 7)$, then we can only realize the generation of an N-quNit singlet state with $N=9$ in an ideal condition.
In addition, If we choose the parameters $(g, \Gamma, \kappa,)/2\pi=(750, 2.3, 3.5)MHz$ which is predicted by Ref.\cite{Spillane}, then the fidelities for the generated singlet states for $N=3,4,5,6$ are about $0.997, 0.993, 0.983, 0.965$, respectively.
Thus, our proposal may be feasible for the generation of singlet states with $N$ limited.

In summary, a novel scheme has been proposed to generate an N-quNit singlet state with adiabatic passage, which may be suitable for arbitrary $N$ in theory.
Although the scale of a singlet state in the present paper is restricted by atomic structures, it may be the first one with cavity QED or similar physical systems.
In addition, quantum information is only stored in ground states, leading the system to nearly be insensitive to atomic spontaneous emission and photon decays.
Finally, the present scheme may be realized with current techniques.

\section{Acknowledgments}

This work is supported by the National Natural Science Foundation of China (Grant nos. 61308012 and 61275215).

\end{document}